\documentclass[fleqn,usenatbib]{mnras}

\usepackage{newtxtext,newtxmath}

\usepackage[T1]{fontenc}
\usepackage{ae,aecompl}

\usepackage{graphicx}
\usepackage{dcolumn}
\usepackage{bm}

\newcommand*{\myfont}{\fontfamily{phv}\selectfont}
\DeclareTextFontCommand{\textmyfont}{\myfont}

\def\ba{\begin{array}}
\def\ea{\end{array}} 
\def\bea{\begin{eqnarray}}
\def\eea{\end{eqnarray}}
\def\beq{\begin{equation}}
\def\eeq{\end{equation}}
\def\ben{\begin{enumerate}}
\def\een{\end{enumerate}}

\def\brr{\begin{array}}
\def\err{\end{array}}

\def\calC{{\S}}
\def\chiC{{\chi_{\S}}}

\title[The Cosmological Constant as a Boundary]{
The Cosmological Constant as a Zero Action Boundary}

\author[Enrique Gazta\~naga]{Enrique Gazta\~naga
\\ 
Institute of Space Sciences (ICE, CSIC), 08193 Barcelona, Spain \\
Institut d\'~Estudis Espacials de Catalunya (IEEC), 08034 Barcelona, Spain
}

\date{Accepted XXX. Received YYY; in original form ZZZ}

\pubyear{2020}

\begin{document}
\label{firstpage}
\pagerange{\pageref{firstpage}--\pageref{lastpage}}
\maketitle

\begin{abstract}
The cosmological constant $\Lambda$ is usually interpreted as Dark Energy (DE) or modified gravity (MG). Here we propose instead that $\Lambda$ corresponds to a boundary term in the action of classical General Relativity. The action is zero for a perfect fluid solution and this fixes  $\Lambda$ to the average density $\rho$ and pressure $p$ inside a primordial causal boundary: $\Lambda = 4\pi G <\rho+3p>$. 
This explains both why the observed value of $\Lambda$ is related to the matter density today and also why other contributions  to $\Lambda$, such as DE or MG, do not produce cosmic expansion.
Cosmic acceleration results from the repulsive boundary force that occurs when the expansion reaches 
the causal horizon. 
This universe is similar to the $\Lambda$CDM universe, except on the largest observable scales, where we expect departures from homogeneity/isotropy, such as CMB anomalies and variations in cosmological parameters indicated by recent observations.
\end{abstract}




\section{Introduction}

Causality is a key property for any physical explanation.  General Relativity (GR) is built as a covariant theory but this does not warrant a causal structure  (see e.g. \citealt{Howard}).  Standard solutions to the Friedmann-Lemaitre-Robertson-Walker (FLRW)
metric lack a proper causal structure: the energy-density and the expansion rate are the same everywhere, in comoving coordinates, no matter how distant. This is not physically possible and has no explanation. 

In the FLRW Universe, the Hubble Horizon $r_H$ is defined as $r_H = c/H(t)$, where $H(t)=\dot{a}/a$ is the  expansion rate 
and $a=a(t)$ is the scale factor. The corresponding comoving coordinate is $\chi_H= r_H/a$.
In normal circumstances $r_H$ has to grow with time because gravity is attractive and slows down the expansion rate $H(t)$.
Scales larger than $r_H$ can't evolve because the time a perturbation takes to travel that distance is shorter than the expansion time. 
This means that $r>r_H$ scales are "frozen out" (structure can not evolve) and are causally disconnected from the rest. 
Thus, $\chi_H$ represents a dynamical causal horizon that is evolving. 

This does not mean that we can't observe scales larger than $\chi_H$. The comoving particle horizon $\chi$, defined as the distance traveled by light since the beginning of time, is the integral of the Hubble Horizon: $\chi = c \int dt/a(t) = \int d\ln{a} ~\chi_H$
and is therefore larger than $\chi_H$.
Thus we can see parts of our Universe which are causally  disconnected from its evolution. This happens with CMB observations, where
the largest angular scales are sometimes called "super-horizon".
In fact, if we look back in time $\chi_H$ becomes smaller and smaller and all observable patches of the Universe become causally disconnected!  This is a very important inconsistency of the FLRW Universe if we seek a causal explanation.

These contradictions can be addressed with inflation. According to inflation \citep{Starobinski1979,Guth1981,Linde1982,Albrecht1982} 
a small primordial (quantum) patch of size  $r_\calC$, which is causally disconnected from the rest of space-time,
starts a DeSitter phase of exponential expansion during which the energy density $\rho_i$ and Hubble rate $H_i$ were constant ($8\pi G \rho_i= 3H_i^2 $). 
The causal horizon $\chi_\S$ is identified with the particle horizon  during inflation: $\chiC = c/(a_i H_i)$
or the Hubble horizon when inflation begins.
This is the largest causally connected scale at the beginning of inflation $a_i$.
This initial patch grows exponentially in physical units, $r_\calC = a\chiC$,
while the corresponding comoving scale $\chiC= c/(a_i H_i)$ and
Hubble horizon $r_H = c/H_i$ remain constant. After 60-70 e-folds $r_\calC$ becomes cosmological in size while $\chi_H$ 
is microscopically small. This is the beginning of the standard FLRW expansion, where the initial Hubble horizon $\chi_H$ 
is negligible as if time had just started. When inflation ends, $\rho_i$ converts into $\rho$ and $p$ (reheating) 
and $\chi_H$ starts to grow inside the  primordial causal horizon $\chiC$.

\begin{figure}
\centering\includegraphics[width=1.0\linewidth]{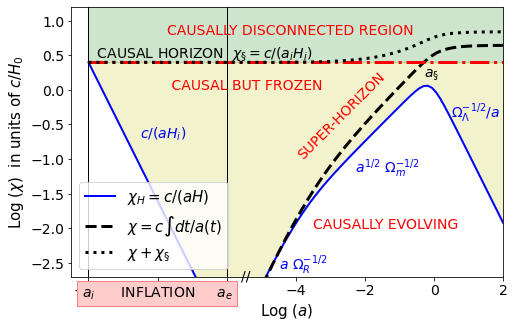}
\caption{Comoving distances  in units of $c/H_0$ as a function of cosmic time $a$ (scale factor). The Hubble horizon  $\chi_H = c/(aH) $, blue continuous line, is compared to the particle horizon $\chi =c \int dt/a(t)$ after inflation, dashed line (dotted line 
includes the $\chiC$ contribution). The  primordial causal horizon $\chiC = c/(a_i H_i)$ (dot-dashed line)  is the particle horizon during inflation.
Larger scales (green shading) are causally disconnected, smaller scales (yellow shading) are dynamically frozen. 
After inflation both $\chi_H$ and $\chi$ grow and by CMB times ($a \simeq 10^{-3}$) we can observe frozen "super-horizon" scales ($\chi>\chi_H$).  At time $a_\calC$ the expansion reaches $\chi_\calC$. This generates a second inflation:  our Universe  is dynamically trapped inside $\chiC$.}
\label{fig:horizon}
\end{figure}

Fig.\ref{fig:horizon} illustrates this causal structure. Models of inflation
expect that $\chiC$ is larger than the particle horizon $\chi_0$ today so that $a_\calC \gg 1$. The second inflation today is then attributed to DE or some other exotic explanation but not to $\chiC$. In our model we will show instead  that $a_\calC \simeq 1$ and $\chi_0 \simeq \chiC$, so the second inflation today is a consequence of the first one. In such scenario, we can observe causally disconnected regions 
($\chi>\chiC$ in Fig.\ref{fig:horizon}) and this should result in cosmic anisotropy and inhomogeneities. Either way, inflation mostly solves the horizon, homogeneity and flatness problems (see e.g. \citealt{Dodelson,Liddle1999}) because we encounter a homogeneous universe that was causally frozen before. 

How can we distinguish if $\Lambda$ is  related to  $\chiC$, to DE or to some other change in GR? Observationally, the boundary  explanation predicts anisotropies and inhomogeneities on scales corresponding to $\chiC$.  Theoretically, we will show that the value of $\Lambda$ is determined by $\chiC$ using the zero action principle. 
The least action principle tells us that arbitrary variations around the true field solutions  produce no changes to the action. 
The action is a functional that assigns a number to each possible metric evolution. The action on-shell (AoS), $S^{on-sh}$, is the value that corresponds to the actual solution of the field equations.
We will show that $S^{on-sh}$ only depends on the value at the boundary of space-time. We are interested here in asymptotically flat (Minkowski) space-time: for a universe of finite age, we expect no effect at scales larger than the true (primordial) dynamical causal horizon (created by inflation).
In such case, we will show that $S^{on-sh}=0$. We call this new property for the Universe the zero action principle. 
This fixes  $\Lambda$ to the average density $\rho$ and pressure $p$ inside the primordial causal boundary set by inflation: $\Lambda = 4\pi G <\rho+3p>$.

To explain cosmic acceleration, scientists have thought of  alternative ways (such as DE, MG or massive gravity) that can contribute to $\Lambda$ or suppress gravity. But each new alternative can't explain why the other contributions can be neglected. The zero action principle implies that other contributions to $\Lambda$ do not produce cosmic expansion because its value is renormalized by the boundary. So the contribution of vacuum energy $\rho_{vac}$ to $\Lambda$ could be as large as the Planck mass ($\rho_{vac} \simeq M_{pl}^4$) and still produce no observational effects. This provides a true solution to the $\Lambda$ problem. 

That $\Lambda$ is related to some cosmic scale, rather than quantum vacuum is hinted by its observed value of $(\Lambda/8\pi G)^{-1/2}$ which is close to  $ \chiC \simeq c/H_0$ (this is sometimes called the coincidence problem).
The holographic principle \citep{Susskind,Maartens} 
provides a way to connect Planck and $\Lambda^{-1/2}$ 
scales via string theories. 
But note how in comoving coordinates $\chiC$ is both scales at the same time! The primordial causal horizon  can start as a Planck scale and become cosmological $c/H_0$ through cosmic expansion \citep{Gaztanaga2020}.

It is generally accepted that GR is not a complete theory, both because of its singularities and also because of its apparent incompatibilities with Quantum Field renormalization theory. 
There are some approaches that relate $\Lambda$ to infrared boundary conditions within Quantum Gravity (see e.g. \citealt{Stenflo,CCB1,CCB2,CCB3,CCB4} and references therein), but they do not seem to bear a direct connection to the classical GR approach presented here.
A more direct connection exist with the discussion in \S16 of \cite{Padmanabhan}, which relates boundary conditions in the action to singularities in the metric.
Here we also find that the  boundary condition in the action (i.e. $\Lambda$) relates to a  singularity in the deSitter metric
(see Appendix \ref{sec:metric}).

Observationally, there is no motivation to modify gravity (MG) on cosmological scales other than to understand $\Lambda$. MG models have been proposed to replace GR using $f(R)$ theories \citep{Nojiri2017}.
These models also require screening mechanisms to avoid Solar System and Astrophysical  tests of gravity \citep{Brax}.
On cosmological scales,  MG models are equivalent to  DE models \citep{Joyce}. But neither approach solves the cosmological constant or coincidence problems. The value of $\Lambda$ is  added and fixed by hand and there is no explanation as to why its value is so small and so close to the matter density today. As we can also add and fix $\Lambda$ by hand in GR, these MG and DE models seem an unnecessary complication (Occam's razor).
But the same principles presented in this work can be easily applied to these GR extensions.

These ideas are similar to the ones in \cite{Gaztanaga2020}, but they are cast here in terms of $S^{on-sh}$. Section \ref{sec:zero} presents the zero action principle. Section \ref{sec:Lambda} introduces $\Lambda$ as a boundary term.
Section \ref{sec:cosmos} explores implications for cosmological observations.
Appendix \ref{sec:metric}  shows how we live inside a Black Hole whose horizon corresponds to $r_\calC$ \citep{hal-03101551}. 
Appendix \ref{sec:PInflation} presents a proposal to start our Universe based on 
$S^{on-sh}=0$.
We use natural units of $c=1$, but show $c$ explicitly when useful. We use sign conventions as  in \cite{Padmanabhan} with Greek (Latin) indexes for 4D (3D).

\section{Zero Action Principle}
\label{sec:zero}

Consider the Einstein-Hilbert action \citep{Hilbert1915}:

\beq
S = \int_M dM \left[ \frac{ R}{16\pi G} +  {\cal L}_m \right],
\label{eq:action}
\eeq
where $dM=\sqrt{-g} d^4 x$ is the invariant volume element, $M$ is the 4D spacetime manifold, $R= R^\mu_\mu = g^{\mu\nu} R_{\mu\nu}$ the Ricci scalar curvature and ${\cal L}_m$ the Lagrangian of the energy-matter content. We can obtain Einstein's field equations for the metric field $g_{\mu\nu}$  from this action by requiring $S$ to be stationary $\delta S=0$ under arbitrary variations of the metric $\delta g^{\mu\nu}$. The solution  is \citep{Einstein1916}:
\beq
  R_{\mu\nu} - \frac{1}{2} g_{\mu\nu} R =  
 8\pi G~T_{\mu\nu} \equiv - \frac{16\pi G}{\sqrt{-g}} \frac{\delta (\sqrt{-g} {\cal L}_m) }{\delta g^{\mu\nu}}.
\label{eq:EFQ}
\eeq
Note that Eq.\ref{eq:EFQ} requires that boundary terms vanish (e.g. see \calC 95 in \citealt{Landau1971}, \citealt{Carroll2004}, \citealt{Padmanabhan}). 
In the notation of \cite{Hawking1996} the boundary term is:
\beq
S = \int_M \left[ \frac{ R}{16\pi G} +  {\cal L}_m \right]
+ \frac{1}{8\pi} \oint_{\partial M} K
\label{eq:actionK}
\eeq
where $K$ is the trace of the extrinsic curvature at the boundary. We can add terms to the action that do not depend on dynamical fields, as they do not change the field equations. As mentioned in the introduction
we are interested in asymptotically flat (Minkowski)  space-times. We can then define the action as:
\beq
S = \int_M \left[ \frac{ R}{16\pi G} +  {\cal L}_m \right]
+ \frac{1}{8\pi} \oint_{\partial M} (K-K_0),
\label{eq:actionK0}
\eeq
where $K_0$ is the trace of the extrinsic curvature of the boundary embedded into asymptotically Minkowski space. This will cancel the boundary term.
We will  show next that the AoS  $S^{on-sh}$ (i.e. the action for the solution of the field equations) for Eq.\ref{eq:action}
is also a boundary term for a perfect fluid. As we require boundary terms to vanish, we have  $S^{on-sh}=0$. We call this the zero action principle.

\subsection{Perfect fluid}

For a perfect isotropic fluid with $\bar{p} =p(t,r)$ isotropic pressure and $\bar{\rho}= \rho(t,r)$  energy-matter density:
\beq
T_{\mu\nu} =  (\bar{\rho}+\bar{p}) u_\mu u_\nu + \bar{p} g_{\mu\nu},
\label{eq:Tmunu}
\eeq 
where $u^\mu= \frac{dx^\mu}{d\tau}$ is the 4-velocity ($u_\mu u^\mu=-1$).
For a generic barotropic fluid with equation of state $w$ and elastic compression $\Pi(\rho)= \int  d\rho p(\rho)/\rho^2 = w \ln{\rho}$ we have
$\bar{p} = w \rho$ and  $\bar{\rho}= \rho [1+w \ln{\rho}]$. This fluid is in general made of several components, each with a different equation of state. 
We describe the metric for an observer comoving with the fluid (so that the spatial velocity is $u^i=0$). We then have:
\bea
T^0_0 =  - \bar{\rho} ~~&;&~~
T^1_1= T^2_2 =T^3_3 =  \bar{p}, \nonumber \\
T &\equiv&  T^\mu_\mu =3 \bar{p}  - \bar{\rho}, 
\eea
and using Eq.\ref{eq:EFQ}:
\bea
R^0_0 =  - 4\pi G (\bar{\rho}  + 3 \bar{p} ) &;&
R^1_1= R^2_2 = R^3_3 = - 4\pi G  (\bar{p}  - \bar{\rho} ), \nonumber \\
R = - 8 \pi G ~T &=& 8 \pi G ( \bar{\rho}- 3 \bar{p}). 
\label{eq:R0}
\eea
The Lagrangian on-shell for a perfect fluid is ${\cal L}_m ^{on-sh} = -\bar{\rho} $ \citep{Minazzoli2012} so that Eq.\ref{eq:action} on-shell becomes:

\beq
S^{on-sh} = - \frac{1}{2} \int_M dM \left[  \bar{\rho} + 3 \bar{p}  \right] =  \int_M dM ~ \frac{ R_0^0 }{8\pi G} \equiv -m/2
\label{eq:action2}
\eeq
where $m$ is the relativistic mass inside the manifold.
Note that this result also holds for anisotropic stress energy with heat flux. In this case we can have different pressure components 
$T_1^1 \neq T_2^2 \neq T_3^3$ and $\bar{p}$ above just corresponds to the mean value.

\subsection{Gaussian flux and Zero Action}
\label{sec:flux}

Consider the geodesic acceleration ${\textmyfont{g}^\mu}=(\textmyfont{g}^0,\textmyfont{g}^i)=(\textmyfont{g}^0,\vec{\textmyfont{g}})$ defined from 
the geodesic deviation equation \citep{Padmanabhan}:
\beq
{\textmyfont{g}^\mu} \equiv
\frac{D^2 \textmyfont{v}^\mu}{D \tau^2} = R^\mu_{\alpha\beta\gamma}   u^\alpha u^\beta \textmyfont{v}^\gamma,
\label{eq:gmu}
\eeq
where $\textmyfont{v}^\mu$ is the separation vector between neighbouring geodesics and $u^\alpha$ is the tangent vector to the geodesic. For an observer following the trajectory of the geodesic $u^\alpha=(1,0)$ and ${\textmyfont{g}^\alpha}=(0,\vec{\textmyfont{g}})$:
\beq
{\textmyfont{g}^i}  = R^i_{00\gamma}  \textmyfont{v}^\gamma
\label{eq:gmu0}.
\eeq
and we can choose the separation vector $v^\mu$ to be the spatial coordinate. The spatial divergence of  $\vec{\textmyfont{g}}$ is then:
\beq
\vec{\nabla} \vec{\textmyfont{g}} = R_0^0 = - 4\pi G (\bar{\rho}  + 3 \bar{p} ),
\label{eq:g}
\eeq
where for the second equality we have used Eq.\ref{eq:R0}.
This equation does not require any approximation and is always valid for a comoving observer
(see Eq.6.105 in  \citealt{Padmanabhan}).
Newtonian gravity is reproduced for the case of non-relativistic matter ($\bar{p} =0$).
The covariant version of Eq.\ref{eq:g} is the relativistic version of Poisson's Equation (see also \citealt{Gaztanaga2020}):
 \beq
\nabla_\mu  {\textmyfont{g}}^\mu  = R_0^0 = - 4\pi G (\bar{\rho}  + 3 \bar{p} ).
\label{eq:g2}
\eeq
The solution to these equations is given by an integral over the usual propagators or retarded  Green functions 
which account for causality and indicates that the AoS has to be estimated on the light-cone.
From Eq.\ref{eq:g2} and Eq.\ref{eq:action2} we can  calculate the value of the AoS using Stoke's theorem to
convert the 4D $M$ integral into a 3D $\partial M$ boundary integral:
\beq
S^{on-sh} = \int_M dM \frac{ R_0^0}{8\pi G} = 
\int_M dM \frac{ \nabla_\mu {\textmyfont{g}^\mu} }{8\pi G}  =
\oint_{\partial M}  \frac{ dV_\mu {\textmyfont{g}^\mu}}{8\pi G}, 
\label{eq:S-on-sh}
\eeq
where $V_\mu$ is a 4D vector whose components are 3D volume elements normal to  $\partial M$
(e.g. $dV_0 \sim dx dy dz$). So  the AoS is just the relativistic version of the Gauss flux, which results in a  boundary term or a total charge. 
Traditionally, we take such boundary terms to be zero at infinity
 (see Eq.4.7.8 in \citealt{Weinberg1972}) because there is no causal connection at infinity. In other words: we expect space-time to be asymptotically Minkowski because particles should be free at spatial infinity for a finite time of evolution. This means that boundary terms should be zero and therefore  $S^{on-sh}=0$.
 
To see this explicitly, consider the boundary $\partial M$ as two spacelike surfaces at $t=t_1$ and $t=t_2$ and one timelike surface at spatial infinity. The usual assumption is that the fields do not contribute at spatial infinity. This is the case for asymptotically flat space-times.
So Eq.\ref{eq:S-on-sh} reduces to two volume $dV_0$ integrals at constant $t$ over $\textmyfont{g}^0$. But $\textmyfont{g}^0=0$ for a comoving observer and we then have $S^{on-sh}=0$.
 
 We also need $S^{on-sh}=0$  because boundary terms must vanish 
 to reproduce Eq.\ref{eq:EFQ} from Eq.\ref{eq:action}.
 In Appendix \ref{sec:AoSTheorem} we present a different derivation for  $S^{on-sh}=0$ as a boundary condition. In Appendix \ref{sec:S-FLRW} we estimate $S^{om-sh}$ explicitly for the FLRW metric and in Appendix \ref{sec:metric} we show how it is possible to 
 combine the FLRW with asymptotically Minkowski boundary conditions.
But note that Eq.\ref{eq:S-on-sh} does not assumed a FLRW metric or homogeneity.

\section{The $\Lambda$ term as a boundary}
\label{sec:Lambda}

As first pointed out by Einstein, we can add a fundamental constant, that here we call  $\Lambda_F$, to the action in Eq.\ref{eq:action} (vacuum energy or Dark Energy are included in the energy content via ${\cal L}_m$ or $T_{\mu\nu}$):
\beq
S = \int_{M} dM \left[ \frac{ R- 2 \Lambda_F}{16\pi G} +  {\cal L}_m \right].
\label{eq:actionC}
\eeq
We can  fulfill the zero action condition $S^{on-sh}=0$ (see above)
by subtracting a constant term ${\cal{B}}$ from the action:
\beq
S = \int_{M} dM \left[ \frac{ R -2 \Lambda_F}{16\pi G} +  {\cal L}_m \right] - {\cal{B}}.
\label{eq:actionC2}
\eeq
This is sometimes called “background subtraction” (another way to write Eq.\ref{eq:actionK0}).
The boundary density ${\cal{B}}/M \equiv  {\Lambda_{\cal{B}}}/{8 \pi G} $ plays the role of a new cosmological constant $\Lambda_{\cal{B}}$, so we can write Eq.\ref{eq:actionC} as:
\beq
S = 
\int_{M} dM \left[ \frac{ R -2 \Lambda }{16\pi G} +  {\cal L}_m, \right],
\label{eq:EHA}
\eeq
where $\Lambda \equiv  \Lambda_F + \Lambda_{\cal{B}}$.
So the value of $\Lambda_F$ is renormalized by the boundary term $\Lambda_{\cal{B}}$, so that its actual value is irrelevant. 
The new field equations are:
\beq
  R_{\mu\nu} - \frac{1}{2} g_{\mu\nu} R + \Lambda g_{\mu\nu} =  
 8\pi G~T_{\mu\nu},
\label{eq:EFQ2}
\eeq
so that  Eq.\ref{eq:R0} transforms into:
\bea
R^0_0 = \Lambda  - 4\pi G (\bar{\rho}  + 3 \bar{p} )  &;&
R^1_1= R^2_2 = R^3_3 = \Lambda- 4\pi G  (\bar{p}  - \bar{\rho} ) \nonumber \\
R = 4\Lambda - 8 \pi G ~T &=& 4\Lambda +8 \pi G ( \bar{\rho}- 3 \bar{p}  ) .
\label{eq:R0Lambda}
\eea
The geodesic equation  Eq.\ref{eq:g2} changes  to:
\beq
\nabla_\mu  {\textmyfont{g}}^\mu  = R_0^0 = \Lambda- 4\pi G (\bar{\rho}  + 3 \bar{p} ).
\label{eq:gLambda}
\eeq
We can see here how  $\Lambda>0$ corresponds to  a repulsive force which opposes gravity, given by the $G$ term. This is the analog of the classical Casimir effect, which induces geometrical forces as a result of a boundary condition (e.g. \citealt{casimir3,casimir2,CASIMIR}).
Replacing  Eq.\ref{eq:R0Lambda} back in the action of Eq.\ref{eq:EHA}, the condition that $S^{on-sh}=0$
fixes $\Lambda$ to:
\beq
\Lambda = 4\pi G
< \bar{\rho} +3 \bar{p}  > ~
\equiv  \frac{4\pi G }{M} \int_{M} dM (\bar{\rho} +3 \bar{p}).
\label{eq:Lambda}
\eeq
Thus $\Lambda = <R_0^0>$ is the action density on-shell (for $\Lambda=0$). As the AoS does not evolve (see Appendix\ref{sec:AoSTheorem}) 
and the primordial causal volume $M_\calC$ is fixed (see  Fig.\ref{fig:horizon} and \S\ref{sec:causal}), $\Lambda$ is also constant.

The condition in Eq.\ref{eq:Lambda}
looks similar to that in \cite{Lombriser2,Lombriser}, who found $\Lambda/4\pi G =<T>=  < 3 \bar{p} - \bar{\rho} >$
(see also 
\citealt{Kaloper2016,Carrol2017}).
Eq.\ref{eq:Lambda} was first obtained by \cite{Gaztanaga2020} starting from the relativistic version of Poisson equation Eq.\ref{eq:gLambda}.  Our new derivation here
shows how $\Lambda$ emerges as a boundary condition to the action and not as a fundamental change to gravity (i.e. $\Lambda_F$ above). A further interpretation as a BH Universe can be seen in \cite{hal-03101551} and Appendix \ref{sec:metric}.

\subsection{Vacuum energy and DE do not gravitate}
\label{sec:Vdoesnot}

It is well known that the effects of $\Lambda$ in Eq.\ref{eq:EFQ2} are equivalent to that of vacuum energy, which has constant energy density and equation of state: $\omega=-1$. 
Replacing $\rho_{vac}$ and $p_{vac}$ in $T_{\mu\nu}$ by
\beq
 \rho_\Lambda \equiv
 \rho_{\rm vac}  + \frac{\Lambda}{8\pi G} 
~~ ; ~~ p_\Lambda = p_{\rm vac}  - \frac{\Lambda}{8\pi G} 
= - \rho_\Lambda,
 \label{eq:rhoHlambda}
\eeq
we can make Eq.\ref{eq:EFQ2} look the same as Eq.\ref{eq:EFQ}.
Equivalently we can  define a new effective energy-momentum tensor $
\tilde{T}_{\mu\nu} \equiv T_{\mu\nu}
-\frac{\Lambda}{8\pi G} g_{\mu\nu}$ \citep{Peebles2003}
so that Eq.\ref{eq:EFQ2} looks the same as Eq.\ref{eq:EFQ}.
So it looks as if  
we could not separate the effects of $\Lambda$  from $\rho_{\rm vac}$. 
But both quantities are related via Eq.\ref{eq:Lambda},
which results in a cancellation of terms in $\rho_\Lambda$.
If we combine $\rho_{\rm vac}$ with matter $\rho_m$ (with $p_m=0$) and radiation $\rho_R$ (with $p_R= \rho_R/3$), Eq.\ref{eq:Lambda} gives:
\beq
\Lambda = 4\pi G \left( <\rho_m + 2\rho_R>- 2\rho_{\rm vac} \right).
\label{eq:m2}
\eeq
so that Eq.\ref{eq:rhoHlambda} results in:
\beq
\rho_{\Lambda}  \equiv \frac{\Lambda}{8\pi G}  + \rho_{\rm vac} =   <\rho_m/2 + \rho_R > 
\label{eq:rhoH2}
\eeq
This shows that vacuum energy cancels out and can not affect the observed value of $\rho_{\Lambda}$ or cosmic expansion. As we will show,
Eq.\ref{eq:rhoH2} also explains 
the coincidence problem \citep{Peebles2003,Carroll2004}
connecting the measured values for  $\rho_m$ and $\rho_\Lambda \simeq 2.3 \rho_m$.

Another way to see this is to combine Eq.\ref{eq:R0Lambda} and Eq.\ref{eq:Lambda}:
\beq
R_0^0 = -4\pi G [(\rho - <\rho>)+ 3(p - <p>) ],
\eeq
which explicitly shows that constant $\rho$ or $p$ components to the energy-momentum do not produce cosmic acceleration, giving $R_0^0=0$
(recall that $R_0^0 = \vec{\nabla} \vec{\textmyfont{g}} = 3\ddot{a}/a$). 
The contribution of DE with equation of state $p_{DE}(a) =\omega ~\rho_{DE}(a)$, is:
\beq
R_0^0 = -4\pi G (1+3\omega)\rho_{DE} [ a^{-3(1+\omega)} - < a^{-3(1+\omega)} > ],
\eeq
where $\rho_{DE}$ is the value today.
So, for $\omega$ close to $\omega\simeq -1$,  DE  produces negligible cosmic acceleration or expansion. 
This removes the original motivation to have DE, as it represents an unnecessary complication of the model (Occam's razor).
Vacuum energy or DE  violate the strong ($\omega>-1/3$) and weak ($\omega>0$) energy conditions in GR (e.g. see \citealt{Visser} and references therein). As shown above, in our interpretation where $\Lambda$ is a boundary term given by  Eq.\ref{eq:Lambda}, vacuum energy does not gravitate and DE is not needed, so these conditions are not necessarily violated. 

In Unimodular Gravity, a modification of classical GR,
$\Lambda$ appears as a Lagrange multiplier or, equivalently, the determinant of the metric is fixed.
In this situation $\Lambda$ does not gravitate (see \citealt{Anderson,Smolin,GarciaBellido:2011de,Percacci} and references therein).
This is different from what we find here, in classical GR, where $\Lambda$ gravitates but it cancels $\rho_{vac}$, DE
or $\Lambda_F$.

\subsection{Causal Boundary and inflation}
\label{sec:causal}

Without inflation the different parts of the observable Universe are causally disconnected.
According to inflation, our Universe is inside a large causal horizon $\chiC$, which could be much larger than our observable Universe today (so that $\chiC \gg \chi_0$).
In such case we expect the flux to be zero at $\chiC$, i.e. $S^{on-sh}=0$ in Eq.\ref{eq:S-on-sh}, because sources produce no flux at distances that are out of causal reach. During inflation, the dynamics of the Universe are frozen out. After inflation our new Hubble horizon $\chi_H$ is very small $\chi_H \ll \chiC$ so we can assume that the Universe inside $\chiC$ follows the (flat) FLRW metric (see Fig.\ref{fig:horizon}). 
For the FLRW Universe,  we have $\rho_\Lambda \simeq \rho_m/(2 <a^3>)$ (see Appendix \ref{sec:S-FLRW}).
The average  $<a^3>$ is over the comoving $\chi$ in the light-cone.
If $\chi$ is large compare to $\chi_0 \equiv \chi(a=1)$ then 
$\rho_\Lambda \rightarrow 0$ as $a \rightarrow \infty$. 
This is contrary to current observations which find: $\rho_\Lambda \simeq 2.3 \rho_m$, which corresponds to $<a^3> \simeq 0.22$ and therefore $\chiC \simeq \chi_0$.
So current measurements of $\rho_\Lambda \neq 0$ are the smoking gun for primordial inflation with $\chiC \simeq \chi_0$.

\section{Late Time Cosmic acceleration}
\label{sec:cosmos}

If we assume that the Universe is isotropic and homogeneous inside $\chiC$ we have the FLRW metric in comoving coordinates $(t,\chi)$:
\beq
ds^2= g_{\mu\nu} dx^\mu dx^\nu = -dt^2 + a(t)^2 \left[ d\chi^2 + S_k^2(\chi) dA^2 \right].
\label{eq:frw}
\eeq
where $dA^2 = cos^2(\theta) d\phi^2 + d\theta^2$ is the solid angle element.
For a closed (or open) Universe of radius $\chi_k$ we have
$S_k=\chi_k \sin[\chi/\chi_k]$ 
(or $\chi_k \sinh[\chi/\chi_k]$). For the flat case $\chi_k \rightarrow \infty$  so that $S_k=\chi$. The FLRW Universe in Eq.\ref{eq:frw}, filled with matter $\rho_m \sim a^{-3}$ ($p_m=0$), radiation $\rho_R \sim a^{-4}$ ($p_R=\rho_R/3$) and $\rho_\Lambda$
has a expansion $H$ and acceleration  rate $q$ (e.g. \citealt{Peebles2003,Weinberg2008}):

\bea
& 3 H^2  = 8\pi G
(\rho_m + \rho_R + \rho_\Lambda)
 -3/(a \chi_k)^2
 \label{eq:Hubble-Law}
 \\ 
& 3q H^2 \equiv 3 \ddot{a}/a 
= R_0^0 = \Lambda -4\pi G(\rho+3p) 
\label{eq:ddota}
\eea
where $ \Omega_X \equiv  \frac{\rho_X}{\rho_c}$
and ${\rho_c} \equiv \frac{3 H^2}{8\pi G}$.
The observed positive value for the cosmic acceleration $q\simeq 0.55$ indicates that $\rho_\Lambda>0$ and corresponds to $
\Omega_\Lambda \simeq 2.3 \Omega_m \simeq  0.7$ today for the flat case $\Omega_k \simeq 0$. These values produce $\chi_H$ as shown numerically in Fig.\ref{fig:horizon}.
The Hubble radius corresponding to the causal horizon
 is $r_\calC = \Omega_\Lambda^{-1/2} \simeq 1.2$, in units of $c/H_0$, which corresponds 
 to a Black Hole with $m \simeq 5.8 \times 10^{22} M_{\odot}$ (see Eq.\ref{eq:BHmass}).
In terms of the FLRW  frame $(t,\chi)$ what is relevant is the corresponding comoving particle horizon coordinate $\chi_\calC$ (see Fig.\ref{fig:horizon}).
We can estimate $\chi_\calC$ from  the average in Eq.\ref{eq:rhoH2} with the information we have from observations after inflation (see Fig.\ref{fig:horizon}). Inflation creates a large causal primordial horizon $\chiC$ inside which the Universe is homogeneous and nearly flat. The particle horizon after inflation is  $\chi$ in Eq.\ref{eq:frw}, $ dt = a(t) d\chi$, so we can write:

\begin{figure}
\centering\includegraphics[width=1.0\linewidth]{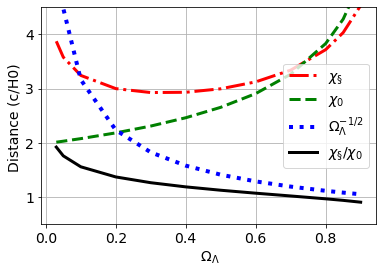}
\caption{Distances (in units of $c/H_0$) as a function of $\Omega_\Lambda$.
The Hubble radius of the causal horizon $r_\calC = \Omega_\Lambda^{-1/2}c/H_0$ (dotted line) is compared with  $\chi_0$, the comoving particle horizon today (dashed line), which increases with $\Omega_\Lambda$.
The comoving primordial causal horizon $\chiC$ (dot-dashed line)  is a combination of the later two and has a minimum size of $\chiC \simeq 3$. The
continuous line shows $\chiC$ in units of $\chi_0$, which is always decreasing. A large boundary, $\chiC>\chi_0$, produces $\Omega_\Lambda \rightarrow 0$, while  $\chiC<\chi_0$ gives $\Omega_\Lambda \rightarrow 1$.}
\label{fig:chiH}
\end{figure}

\beq
\chi(a) =  \int_{t_e}^{t}  \frac{dt}{a(t)} = \int_{a_e}^{a}  \frac{ d\ln a}{a H(a)} = \chi_0 - \bar{\chi}(a),
\label{eq:chia}
\eeq
where $t_e$ and $a_e$ represent the time and scale factor when inflation
ends. The particle horizon today is $\chi_0 \equiv \chi(a=1) \simeq 3\chi_H$ and $\bar{\chi}(a) = \int_a^1 da/(a^2H)$ is the radial lookback time coordinate so that
$d_A(\bar{\chi})=S_k(\bar{\chi})$ is the comoving angular diameter distance. 
We can now estimate the average in 
Eq.\ref{eq:rhoH2} in the light-cone to $\chi_{\calC}$:

\beq
\rho_{\Lambda}   = 
~ \frac{\int_{0}^{\chi_{\calC}}  d\chi ~ S_k^2(\chi)~(1-\cos{\frac{\chiC}{\chi}}) ~{a^3}~ (\rho_m a^{-3} + 2\rho_R a^{-4})~ }
{2\int_{0}^{\chi_{\calC}}  d\chi  ~S_k^2(\chi)~(1-\cos{\frac{\chiC}{\chi}})  ~a^3 } ,
\label{Eq:rhoHchi}
\eeq
where $a=a(\chi)$ is the light-cone crossing corresponding to the inverse of $\chi(a)$ in Eq.\ref{eq:chia}, which also depends on $\rho_\Lambda$ via the Hubble rate $H$. The term: $(1-cos {\frac{\chiC}{\chi}})$ comes from the limits of $\chiC$ on the solid angle $dA$ and was not included in  \cite{Gaztanaga2020}, which explains the small differences in the final results.
To do this integral, it is easier to change variables to $a$:
\beq
\Omega_{\Lambda}   = 
~ \frac{\int_{0}^{ a_\calC}  \frac{ada}{H} ~ S_k^2(\chi)~(1-\cos{\frac{\chiC}{\chi}})   ~(\Omega_m a^{-3} + 2\Omega_R a^{-4})~ }
{2 \int_{0}^{ a_\calC}  \frac{ada}{H} ~ S_k^2(\chi)~(1-\cos{\frac{\chiC}{\chi}}) } ,
\label{Eq:rhoHchi_a}
\eeq
where $\chi=\chi(a)$ in Eq.\ref{eq:chia} and $a_\calC \equiv a(\chiC)$.
We have  divided by $\rho_c$ to express the result in terms of adimensional ratios $\Omega$. For each value of $\chiC$ (and $\Omega_k$, with $\Omega_R \simeq$4.2E-5)
we can find numerically the corresponding $\Omega_\Lambda$. The result is shown in Fig.\ref{fig:chiH}.
For $\Omega_\Lambda
\simeq 0.7 \pm 0.1$ and $\Omega_k \simeq 0$ we find:

\begin{figure}
\centering\includegraphics[width=1.0\linewidth]{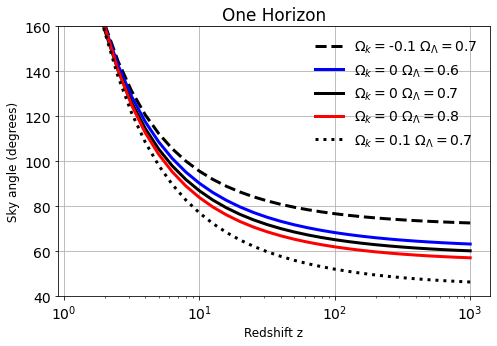}
\caption{ Angle on the sky 
$\theta_\calC(z)=\chi_\calC/d_A(z)$ 
corresponding to a (transverse) primordial causal boundary $\chi_\calC$, for different $\Omega_k$ and $\Omega_\Lambda$. At $z\simeq 2$ about half of the sky is causally disconnected. The lack of structure at $\theta>60$ deg.
 in the CMB ($z \simeq 10^3$)  favours $\Omega_k \lessapprox 0$  and $\Omega_\Lambda \lessapprox 0.7$. We should look for other tracers of $\chiC$ at lower $z$. }
\label{fig:angle}
\end{figure}

\beq
\chiC = \left( 3.34 \pm 0.18 \right) 
\frac{c}{H_0} 
~~~\rm{and}~~~ a_\calC \equiv a(\chiC) \simeq 1.08 \pm 0.16 ,
\label{eq:chi_H}
\eeq
compared to: $\chi_0=3.26 c/H_0$. 
Because $\chi_{\calC}< \pi \chi_0$ we can actually measure 
$\chi_{\calC}$ on the sky. 
At the CMB, $\chiC$ corresponds to an angle:
$
\theta_\calC(z_{\rm{CMB}}) \equiv \chiC/d_A
\simeq 60$ deg.
This  boundary could therefore result in
CMB anomalies or tensions in measurements. 
Because $\chi_{\calC}$ in Eq.\ref{eq:chi_H} is closed to $\chi_0$, we could wonder if inflation lasted enough time to make $\Omega_k \simeq 0$ today. In fact,  $\Omega_k<0$ is not ruled out by observations \citep{DiValentino}.
For $\Omega_k= -0.06$ and $\Omega_\Lambda \simeq 0.7 \pm 0.1$ we find $\chi_{\calC} = ( 3.39 \pm 0.27 )\frac{c}{H_0}$ compared to $\chi_0=3.31$ and $\theta_\calC(z_{\rm{CMB}}) \simeq 68$ deg. Fig.\ref{fig:angle}  shows the angle on the sky $\theta_\calC(z)=\chi_\calC/d_A(z)$ corresponding to a transverse $\chiC$.

Because $d_A(z_{CMB}) \simeq \chiC$,  we expect the cosmological parameters in our local universe to be slightly different from the ones that we see in the CMB, as they come from different primordial causal patches. This could explain tensions in measurements of $H_0$ and other cosmological parameters. The full CMB sky covers
transverse scales up to $\simeq \pi d_A \simeq \pi \chiC$, so that we are able to measure a few separate causal horizons on the CMB sky \citep{FG20}.

\section{Discussion and Conclusions}

We have argued that our universe not only has "zero energy" (i.e. critical density) but also zero AoS.  This generates a boundary term for the  action which fixes the cosmological constant to  $\Lambda=4\pi G <\rho+3p>$,  i.e. Eq.\ref{eq:Lambda}. As a consequence vacuum energy, DE or a fundamental $\Lambda_F$ do not gravitate (see \S\ref{sec:Vdoesnot}). This provides a true solution to the cosmological problem. It predicts that the observed value of $\rho_\Lambda$ is related to  the mean matter and radiation content inside the boundary $\chi<\chiC$, see Eq.\ref{eq:rhoH2} and \ref{Eq:rhoHchi}. In our interpretation,
$\Lambda$ is not a dark energy component or a fundamental constant $\Lambda_F$ \citep{Weinberg1989,Huterer,Carroll1992}, but just a boundary term (see \S\ref{sec:Lambda}).
 Measurements of  $\rho_\Lambda$ allow us to estimate  $\chiC$ (see Eq.\ref{eq:chi_H} and Fig.\ref{fig:chiH}).
For $\chiC > \chi_0$ we have  $\Omega_\Lambda \simeq 0$, which is contrary to observations. While $\Omega_\Lambda \simeq 1$ indicates that  $\chiC \simeq \chi_0$. This can be used to constrain models of inflation
\citep{Gaztanaga2020}. 
In contrast to the Hubble horizon, $\chi_H = c/(aH) $, which increases with $a$, the primordial boundary $\chiC$ remains constant throughout cosmic evolution (see Fig.\ref{fig:horizon}).
The $\Lambda$ boundary corresponds to a singularity in the metric of our Universe. We are dynamically trapped inside $\chiC$,  like observers inside a Black Hole (BH). In Appendix \ref{sec:metric}) we show how $\chiC$ corresponds
to the BH horizon, which explains how we can have an asymptotically Minkowski boundary and a FLRW metric at the same time \citep{hal-03101551}.
Cosmic acceleration occurs when our particle horizon approaches this primordial boundary. Thus late time cosmic acceleration is the smoking gun of primordial inflation.

\begin{figure}
\centering\includegraphics[width=1.0\linewidth]
{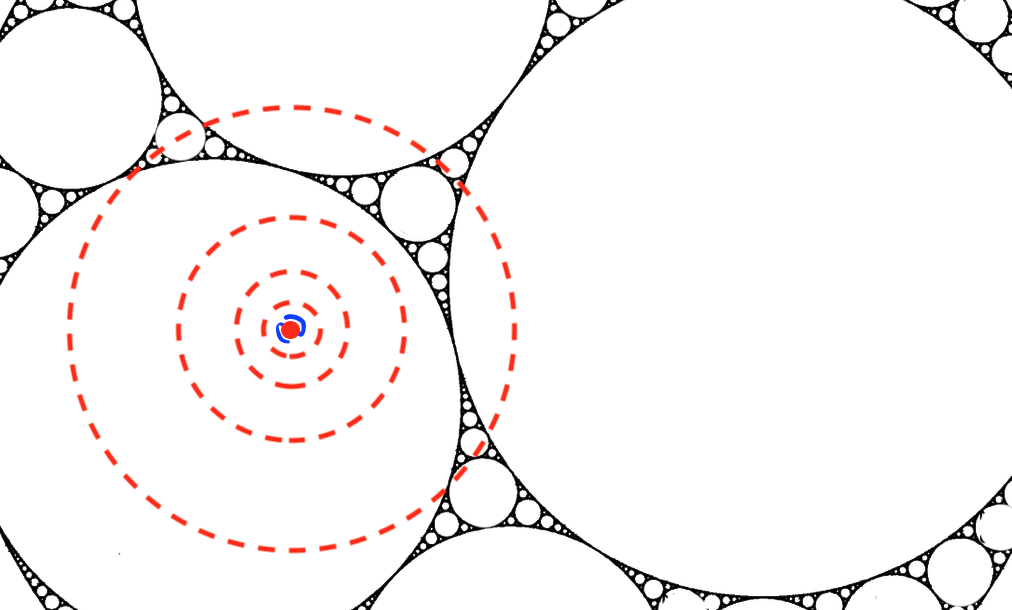}
\caption{Illustration of the
causal structure of the Universe after inflation.
Each circle (with continuous lines) represent an independent Universe: a causally disconnected horizon with different cosmological parameters and values of $\chiC$ (and $\rho_\Lambda$). 
Our galaxy is depicted at the center of the dashed concentric circles which represent the growth evolution of our particle horizon $\chi_0$ after inflation ends.
A maximum size is reached when $\chi_0 \simeq \chiC$.  Further evolution freezes out as our Hubble horizon shrinks again because  the expansion becomes dominated by  $\rho_\Lambda$ (see Fig.\ref{fig:horizon}).}
\label{fig:apollonian}
\end{figure}

Each causally disconnected patch (or horizon) that emerges from inflation evolves like a separate Universe (see Fig.\ref{fig:apollonian} and Appendix \ref{sec:PInflation}).
Our past evolution is not influenced
by other horizons because $\chi_H<\chiC$.
This is why the zero action principle applies to each separate horizon.
But we have a window to observe other horizons because our observable Universe (or past particle horizon) in Eq.\ref{eq:chia} is larger than the Hubble horizon: $\chi_0 \simeq 3\chi_H$  (see dashed line in  Fig.\ref{fig:horizon}) and $\chi_0 \simeq \chiC$ so that $\pi \chi_0 \simeq \pi d_A > \chiC$.
Properties might not be homogeneous across nearby primordial patches which could result in cosmic anisotropies. One could imagine that this could lead to large scale super-horizon fluctuations, like in the Grishchuk-Zel'dovich effect on the CMB quadrupole \citep{Grishchuk}. 
But recall from Fig.\ref{fig:horizon} that the so called super-horizon scales are frozen scales smaller than $\chiC$, the horizon during inflation. Each causal horizon is by definition independent  so the spectrum of fluctuations  vanish on causal horizon scales. This is in agreement with the measured CMB quadrupole, which is smaller, and not larger, than expected. 
Solutions in separate primordial horizons should be matched into a larger manifold solution \citep{Sanghai-Clifton,hal-03101551}.
Continuity across contiguous regions then forces any uncorrelated differences to be small. This could explain why the mean CMB temperature can be similar across nearby disconnected patches in the sky.

That the primordial causal boundary scale, $\chiC$, is similar in size  to our observable Universe today, $\chi_0$, solves the  coincidence problem \citep{Peebles2003,Carroll2004} of why is $\rho_\Lambda \simeq 2.3 \rho_m$ or why $(\Lambda/8\pi G)^{-1/2}$ is close to $c/H_0$. But it rises new questions: why $\chiC \simeq \chi_0$?
Fig.\ref{fig:chiH} shows how very different values of $\Omega_\Lambda$ have similar
 $\chiC/\chi_0$ values. So may be this is not such a big coincidence.
In terms of anthropic reasoning, 
at earlier times the Universe is dominated by radiation and there are no stars or galaxies to host observers. 
Moreover, $\chi_0 \simeq \chiC$, has the largest possible Hubble radius (see Fig.\ref{fig:horizon}) with the highest chances to host observers like us.  Ultimately, the reason for   $\chiC \simeq \chi_0$ resides in the details of the initial conditions (see also Appendix \ref{sec:PInflation}).

We should be able to find observational evidence of such a primordial boundary $\chiC$ in our sky (see Fig.\ref{fig:angle}). Because separate horizons are not causally connected, we  expect to see no statistical correlations in the sky on angular scales  $\theta > \theta_\calC$. This seems in agreement with the anomalous 
lack of correlations (respect to LCDM model) in the CMB temperature correlations $w(\theta)$ for $\theta \gtrapprox\ 60$deg. Such anomaly has been used to predict $\Omega_\Lambda \simeq 0.7$ for a flat Universe with independence of any other measurements \citep{Gaztanaga2020}. 
But note that the value of $\Omega_\Lambda$ that we measure locally, could be different from the one affecting the faraway distance where CMB photons were emitted.
Tensions between measurements of cosmological parameters (or fundamental constants) from very different redshifts  or different regions of the sky (at high $z>2$) could be related to such in-homogeneities. 

In a recent analysis \cite{FG20} found strong evidence for such horizons
in the coherent variation of cosmological parameter  in the CMB sky. 
We should look closer for such differences  to further validate or falsify these ideas.

\section*{Acknowledgments}
We thank  C.M.Baugh, B.Camacho, E.Elizalde, P.Fosalba, J.Garcia-Bellido, L.Hui, L.Lombriser, J.Peebles, P.Renard and I.Tutusaus for their feedback. This work has been supported by spanish MINECO  grants PGC2018-102021-B-100 and EU grants LACEGAL
734374 and EWC 776247 with ERDF funds.
IEEC is funded by the CERCA program of the Generalitat de Catalunya.

\section*{Data Availability Statement}

The data and codes used in this article will be shared on request.

\bibliographystyle{mnras}
\bibliography{gaztanaga} 

\appendix

\section{AoS Theorem}
\label{sec:AoSTheorem}

We next enunciate and proof what we call the Action on-shell (AoS) Theorem, which basically says that the AoS is fixed by its boundary. 
This is reminiscent of the holographic principle \citep{Susskind,Maartens}. Indeed the  Einstein–Hilbert action in Eq.\ref{eq:action} possesses such holographic relation between the surface and bulk terms (see Eq.15.45 in \citealt{Padmanabhan}). 

This theorem applies to any stationary action principle. 
But we note that the proof presented here lacks the mathematical rigor of the Stokes' Theorem which was used to derive the zero action principle in section \ref{sec:flux}.
Our arguments below assume continuous and smooth functions and do not account for situations with singularities. 

{\noindent \bf Definitions:}  given an action functional:

\beq
S[g,M] = \int_M dM ~ {\cal L}_m[g] ,
\label{eq:General-action}
\eeq
where $M$ is a spacetime manifold and 
$dM= \sqrt{-g} d^4 x$ is the invariant 4D volume and
$g=g(x)$ stands for any dynamical field in our problem, including the metric $g^{\mu\nu}$.
Assume  that $S$ obeys an stationary action principle, i.e. the equations of motion for $g$ are such that arbitrary small variation $\delta g$ around the solution $g^*$ produce no changes in $S$, so that:  
\beq
\delta  S[g^*, M]\equiv S[g^*+\delta g, M] -S[g^*, M]
=0.
\label{eq:actionP}
\eeq
The AoS is defined as $S^{on-sh} \equiv S[g^*,M]$.

{\noindent \bf Theorem:}  $S^{on-sh} =  S[g^*, \partial M]$ where  $\partial M$ is a boundary of $M$.

{\noindent \bf Proof:} Consider a partition of $M = M_1 +M_2$:
\beq
S[g^*, M] = S[g^*, M_1] + S[g^*, M_2],
\label{eq:M1M2}
\eeq
it follows from Eq.\ref{eq:actionP}:
\beq
S[g^*, M] = 
S[g^*+\delta g, M_1] + S[g^*+\delta g, M_2].
\eeq
Because $\delta g$ can be arbitrary we can choose:
\beq
\delta g  = \left\{ \begin{array}{ll} 
 \delta g_1 & {\text{for}} ~ ~ M_1 \\
0 &  {\text{for}} ~ ~ M_2 \\
\end{array} \right.
\label{s(k)2}
\eeq
 so that:
\beq
S[g^*, M] = 
S[g^*+\delta g_1, M_1] + S[g^*, M_2] ,
\eeq
and comparing with Eq.\ref{eq:M1M2} we find:
\beq
S[g^*, M_1] = S[g^*+\delta g_1, M_1] ,
\eeq
so that the action is also stationary as in
Eq.\ref{eq:actionP} for any part of $M$.
Consider a volume element $dM$ centered around  a coordinate $(t,x)$ of the manifold. We can choose $M_1=dM(t,x)$ so that each step in the integration in Eq.\ref{eq:General-action}
is stationary:
\beq
\delta S[g^*, dM(t,x)] =0  .
\label{eq:dM}
\eeq
The solution $g^*(t,x)$ evolves when we move one step $dM$ away from $(t,x)$. We denote this 
$\delta g^*(t,x)$ to distinguish it from some arbitrary variation of the field $\delta g$.
Because the action for each volume element $dM(t,x)$ is stationary  around an arbitrary variation of $\delta g$, it should also be stationary if we choose $\delta g = \delta g^*(t,x)$, so that:
 \beq
 S[g^*, dM(t,x)] =  S[g^*+ \delta g^*, dM(t,x)],
\label{eq:dM2}
\eeq
which shows that the  evolution of the field in step $dM$ does not change the action. As this is true for each step in the action integral, we can conclude that the  action integral value on-shell is the same as the one in its boundary: $S[g^*, M] =  S[g^*, \partial M]$.
Q.E.D.

{\noindent \bf Corollary-I:} If the Universe has a  asymptotically Minkowski boundary, the AoS must be zero, $S^{on-sh}=0$, for its whole evolution.

{\noindent \bf Proof:}  Because the Minkowski metric has $S^{on-sh}=0$, we can use the AoS Theorem to conclude that $S^{on-sh}=0$ also for the full evolution. Q.E.D. 

An example of Corollary-I is the zero Gaussian flux condition presented in section \ref{sec:flux}. Another example is the
 Schwarzschild metric, which asymptotically goes to Minkowski and produces $S^{on-sh}=0$ (if we avoid the singularity at $r=0$).
Other examples where $S^{on-sh}=0$ are the Dirac action or the harmonic oscillator action. More generally, the kinetic term of a field theory ${\cal L}_m \propto \psi_i L_{ij} \psi_j$ (where $L_{ij}$ is a linear operator) is zero on-shell because the equation of motion is $L_{ij} \psi_j = 0$. Potential terms are usually assumed to be zero at the boundaries because of causality.

{\noindent \bf Corollary-II:} If the Universe started out of nothing, the AoS must be zero, $S^{on-sh}=0$, for its whole evolution.

{\noindent \bf Proof:} The solution for the field equation for a Universe without content  $T_{\mu\nu}=0$ is the Minkowski metric. Using Corollary-I we 
conclude that $S^{on-sh}=0$ also for the full evolution. Q.E.D.

\section{The Action for the FLRW metric}
\label{sec:S-FLRW}

To estimate  Eq.\ref{eq:action2} for a flat matter dominated FLRW metric, we choose our location as the origin of the radial coordinate and the end of inflation as the initial  condition. 
For simplicity consider the flat matter dominated Universe
in the FLRW metric in Eq.\ref{eq:frw} and Eq.\ref{eq:Hubble-Law}. 
The 4D volume element is $dM = dt \sqrt{-g} d^3x = dt  ~ a^3 4\pi \chi^2 d\chi$ where we have used: $\sqrt{-g}= a^3$. 
As we integrate in the radial direction we need to update time to the light cone $dt=d\chi/a$ because the solution of the field equations obey causality (on-shell). The integral them becomes integral over $\chi=\chi(a)$ or $a=a(\chi)$ (see Eq.\ref{eq:chia}):
\beq
S^{on-sh}_{FLRW} = -\frac{1}{2}
\int_{0}^{\chi}  d\chi ~ 4\pi \chi^2 ~{a^3}~(\rho_m a^{-3})= - V ~\frac{\rho_m}{2}.
\label{eq:action-frw}
\eeq
where $V= 4/3 \pi \chi^3$ is the 3D comoving volume and $\rho_m$ the matter density today ($a=1$). 
This integral is unbounded and diverges as we increase the comoving volume $V$ (or the time involved). This illustrates the issue with the FLRW metric that we mentioned in the introduction. It produces boundary terms
because it has a non compact geometry. This is an issue because we can not reproduce Eq.\ref{eq:EFQ} from Eq.\ref{eq:action}, as boundary terms don't cancel out.
In the case of a FLRW model with $\Lambda$, we can use Eq.\ref{eq:EHA} to find:
\bea
S^{on-sh}_{FLRW} &=&   V ~(\frac{\Lambda  <a^3>}{8\pi G} -\frac{\rho_m}{2} ) \\
<a^3> &\equiv& \frac{1}{V}  \int_{0}^{\chi}  d\chi ~ 4\pi \chi^2 ~{a^3}
\label{eq:action-frw2}
\eea
We can see here how we can fix $\Lambda=4\pi G \rho_m/<a^3>$
to cancel this boundary term. 
 Note that in this case $V$ is finite because there is a maximun value for $\chi$ (see Fig.\ref{fig:horizon}). If we also add $\rho_{vac}$ we find: $\rho_\Lambda=\rho_m/(2<a^3>)$ which cancels $\rho_{vac}$ (see \S\ref{sec:Vdoesnot}).

\section{The Black Hole Universe (BHU)}
\label{sec:metric}

\begin{figure}
\centering\includegraphics[width=1.0\linewidth]{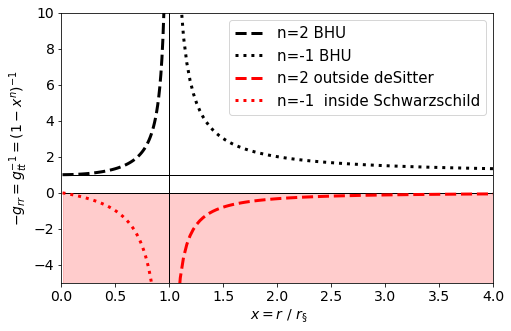}
\caption{Metric elements $g_{tt}^{-1}= -g_{rr} =  (1-x^n)^{-1}$ 
of the Black Hole Universe (BHU) in Eq.\ref{eq:BHU} as a function of $x=r/r_\calC$.
The BHU metric corresponds to Schwarzschild ($n=-1$) for the outside $x>1$  and to deSitter  ($n=2$) for the inside $x<1$, which preserves the signature for all $r$.
The outside of deSitter metric or the inside of Schwarzschild have negative metric signature (red shading) which indicates that they are not causally connected.}
\label{fig:BHU}
\end{figure}

The idea of the causal boundary $\chiC$ results in a non-homogeneous universe. This makes sense for a universe of finite age because there are causally disconnected regions with different densities (see  Fig.\ref{fig:apollonian}).
For $\chi < \chiC$ the FLRW metric should be a good approximation, but we want an asymptotically flat metric on larger scale. How can we combine these two different metrics? 

Consider the flat FLRW metric of Eq.\ref{eq:frw} in comoving coordinates $(t,\chi)$. When $\rho_\Lambda$ dominates the expansion rate: $H^2 \equiv (\dot{a}/a)^2= 8\pi G \rho_\Lambda/3$, which results in exponential inflation: $a(t) \propto e^{Ht}$.
We can change variables from comoving $\chi$ to proper $r=a(t) \chi$ distances, and introduce a new time variable $T=T(t,\chi)$ such that $\partial_t T = ( 1 -H^2 r^2)^{-1}$
and  $\partial_\chi T =  a H r ~ \partial_t T$. In these new variables $(T,r)$, the FLRW metric becomes DeSitter:
\beq
ds^2= -(1- H^2 r^2)~dT^2 + \frac{dr^2}{1- H^2r^2} + r^2 dA^2,
\label{eq:dS}
\eeq
where $H^2= 8\pi G \rho_\Lambda/3$ is constant. So in $(T,r)$ coordinates the FLRW expansion is static (see \citealt{Mitra2012}). DeSitter  metric above is very similar to the Schwarzschild metric:
\beq
ds^2= -(1- 2Gm/r)~dT^2 + \frac{dr^2}{1- 2Gm/r} + r^2 dA^2,
\label{eq:Sch}
\eeq
with an Event Horizon (EH) at $r_\calC= 2Gm$. 
Both metrics are static. DeSitter metric also has an EH at
$r_\calC = 1/H$.
But DeSitter solution corresponds to the interior $r<r_{EH}$, while the Schwarzschild metric represents the exterior $r>r_{EH}$. 
We can match both solutions into a Black Hole Universe (BHU) where we consider DeSitter metric as the interior of the  Schwarzschild metric:
\bea
ds^2 &=& -[1-(r/r_\calC)^n]~dT^2 + \frac{dr^2}{1- (r/r_\calC)^n} + r^2 dA^2, \nonumber \\
n &=& \left\{ \begin{array}{ll} 
 2 & {\text{for}} ~ ~r<r_\calC \\
-1 &  {\text{for}} ~ ~ r>r_\calC \\
\end{array} \right.
\label{eq:BHU}
\eea
which is continuous and smoothed everywhere except at the EH
(see Fig.\ref{fig:BHU}).
This results in $r_\calC=1/H= 2 mG$, which gives:
\bea
\rho_\Lambda &=& 3m/(4\pi r_{\calC}^3),
\label{eq:rhoEH} \\
m &=&  \frac{4\pi}{3} r^3_\calC  <\rho_m/2 + \rho_R >. 
\label{eq:BHmass}
\eea
where in the second equation we have use Eq.\ref{eq:rhoH2}.
The expansion of our universe is not always dominated by $\rho_\Lambda$. But Eq.\ref{eq:dS} is still the exact transformation of the FLRW metric in the Schwarzschild frame  $(T,r)$, by just replacing $H$ by $H(t)$  \citep{hal-03101551}.
Thus, the FLRW metric also has an EH at $r_H \equiv 1/H(t)<r_\calC$. But this is an expanding horizon that grows inside $r_\calC$. As $r_H$ approaches $r_\calC$ a repulsive force (from the $\Lambda$ term or the EH) forces  $r_H=r_\calC$ (see Fig.\ref{fig:horizon}). So Eq.\ref{eq:BHU}-\ref{eq:rhoEH} is valid throughout cosmic evolution (for more details on the BHU see \citealt{hal-03101551}). 

In summary, we can combine the FLRW metric and the Schwarzschild metric into a BHU
to obtain the causal structure required by the zero action principle: $S^{on-sh}=0$.

\section{The Apollonial Multiverse}
\label{sec:PInflation}

The zero action principle could also be used to build a multiverse of BHU.
We can picture the start of our universe as an empty Minkowski space-time with quantum fluctuations of some field
 $\rho_{I}(a)$  with equation of state $\omega$ or just false vacuum BH \citep{hal-03101551}.
Because $S^{on-sh}=0$, this fluctuation will generate a boundary $\Lambda$ term (also call a BH).
We can consider these fluctuations homogeneous inside some (quantum) causal scale $a_i\chiC$. The Hubble rate is then:
\beq
H^2 = \frac{8\pi G}{3} \left[ \rho_\Lambda + \rho_{I} ~(a/a_i)^{-3(1+\omega)} \right],
\eeq
where $a_i$ is some reference primordial time and
from Eq.\ref{Eq:rhoHchi}:
\beq
\rho_{\Lambda}   =  \rho_I \frac{1+3\omega}{2} 
~ \frac{\int_{0}^{\chi_{\calC}}  d\chi ~ S_k^2(\chi) ~{a^3}~( (a/a_i)^{-3(1+\omega)} )~ }
{\int_{0}^{\chi_{\calC}}  d\chi  ~S_k^2(\chi)  ~a^3 },
 \label{eq:rhoH3b}
\eeq
where $\chiC$ (and $a_\calC \equiv a(\chiC)$)  corresponds to the causal scale. 
For $a_\calC<a_i$ and $\omega >-1$ (or $a_\calC>a_i$ and $\omega <-1$)
the $\rho_{\Lambda}$ term  dominates over $\rho_I ~(a/a_i)^{-3(1+\omega)}$. This will generate an inflationary expansion. The quantum fluctuation will be enlarged and become macroscopic. This inflation will also make the Hubble horizon
smaller than  $\chiC$ (see Fig.\ref{fig:horizon}) so our observable Universe will eventually become quite large but trapped inside $\chiC$.
This will  happen in different locations of the  initial Minkowski manifold so that we can end up with multiple disconnected and inflating Universes. Each Universe will have different $\chiC$ and $\rho_\Lambda$ and they will inflate until they touch each other creating an Apollonial fractal, as illustrated by Fig.\ref{fig:apollonian}.

Another way to picture this, for an external observer, is that we will have a network of interacting BHU (see Appendix \S\ref{sec:metric}) which could also be in equilibrium with some external (Hawking) radiation or matter fields.
It seems reasonable to predict that their interactions could end inflation 
inside some BHU (e.g. because of the dynamical collisions) or just because false vacuum decay or false vacuum rolling \citep{hal-03101551}. The dynamical energy of the expansion
could transform into thermal energy, in a similar way to the reheating mechanism in traditional inflationary models.  

Inflation will happen under quite  general considerations according to the zero action principle.
The principle will still apply to each separate Universe (with some influence of its surroundings)  because the interaction with nearby horizons will be limited in time and space.  
Later quantum fluctuations within one universe represent negligible contribution and do not change $\chiC$ or $H$. 

Our universe is now entering a second inflationary phase and it is possible that we could cycle to a situation similar to our previous inflation. So repeated phases of inflation could also play a role to understand our Universe and the value of the cosmological parameters. Could this result in an eternal return? or in natural selection \citep{Smolin1992}?
Details of this model need to be worked out.

\label{lastpage}
\end{document}